# Denoising Signals in Cognitive Radio Systems Using An Evolutionary Algorithm Based Adaptive Filter

Adnan Quadri, Mohsen Riahi Manesh and Naima Kaabouch
Department of Electrical Engineering,
University of North Dakota, Grand Forks, 58203, USA
E-Mail: adnan.quadri@und.edu; mohsen.riahimanesh@und.edu; naima.kaabouch@engr.und.edu

*Abstract*— Noise originating from several sources in a RF environment degrades the performance of communication systems. In wideband systems, such as cognitive radios, noise at the receiver can originate from non-linearity present in the RF front end, time-varying thermal noise within the receiver radio system, and noise from adjacent network nodes. Several denoising techniques have been proposed for cognitive radios, some of which are applied during spectrum sensing and others to received noisy signal during communication. Examples of some of these techniques used for noise cancellation in received signals are least mean square (LMS) and its variants. However, these algorithms have low performance with non-linear signals and cannot locate a global optimum solution for noise cancellation. Therefore, application of global search optimization techniques, such as evolutionary algorithms, is considered for noise cancellation. In this paper, particle swarm optimization (PSO) and LMS algorithms are implemented and their performances are evaluated. Extensive simulations were performed where Gaussian and non-linear random noise were added to the transmitted signal. The performance comparison was done using two metrics: bit error rate and mean square error. The results show that PSO outperforms LMS under both Gaussian and non-linear random noise.

## I. INTRODUCTION

One of the challenges for communication systems is noise, which degrades the performance of data transmission between a transmitter and a receiver. Examples of sources of noise include: non-linearity present in the RF front end, time-varying thermal noise within the receiver radio system, and noise from adjacent network nodes or RF environment [1-3]. In addition, crosstalk, shadowing, and path loss are also factors that impact the integrity of signals [3]. To deal with the noise, conventional communication systems employ fixed hardware [4], which limits performance and lacks dynamic functionalities. However, software based systems enable reconfigurability by utilizing multi-purpose digital programmable devices, such as FPGAs, for signal processing instead of specifically designed hardware [5].

Example of such reconfigurable and adaptive technologies is Cognitive Radio (CR). CR systems built on software defined radio (SDR) are wideband transceivers operating with full-duplex communication. In addition to the previously mentioned sources of noise, CR systems are impacted by several non-linear system-induced noise as CR has to perform multiple advanced and complex signal processing operations over a wide range of frequency bands. Noise in CR can be generated from the interference caused by multiple bands during spectrum sensing, noise saturation of the CR receiver by the co-located CR transmitter operating at the same time and frequency band during full-duplex communication, and system non-linearity [6].

To mitigate the impact of noise in a reconfigurable system such as CR, adaptive filters based denoising techniques enable readjusting filter parameters according to the channel and signals. In general, adaptive filter based denoising techniques use algorithms that can be classified in two categories: gradient-descent and non-gradient based algorithms. Gradient-descent based techniques are multivariate optimization techniques that start with an assigned initial value and follows the negative of gradient to reach the desired local minimum. Examples of these techniques include least mean square (LMS) and its variants -- normalized LMS (NLMS) [7], recursive least square (RLS) [8], and filtered x-LMS (FxLMS) [9]. These algorithms find optimal weight solution for the adaptive filter to minimize error and cancel residual noise. However, as these techniques perform local optimizations, they can only locate local minima, thus failing to find a global optimum solution to minimize the error signal. In addition, these techniques are also dependent on step size based initiation. A high step size value degrades the steady state of the filter, whereas a low step size value delays the convergence of the filter [10]. Moreover, LMS algorithms were found to perform poorly with colored noise and in non-linear environments [10].

To overcome the problem of locating global minima of an error surface, non-gradient algorithms, also known as global search optimization techniques, can be applied. Examples of such algorithms include genetic, artificial bee colony (ABC), cuckoo search, and particle swarm optimization algorithms. Some of these algorithms, such as the genetic algorithm, require selecting appropriate initialization values for the process of mutation and crossover to converge at a steady rate [11]. Often, the selection of appropriate values for this initialization of variables is found to be case-dependent and estimated through empirical observations. Several other research works proposed improved version of these algorithms by applying self-adaptive methods of defining the initialization variables [11 - 13]. PSO algorithm, on the other hand, does not rely on a specific single variable initialization, such as the step size in gradient algorithms and is less complex [14].

To the best of our knowledge, the prospect of using evolutionary algorithm based adaptive filters, specifically for CR systems, has not yet been explored. But some research works proposed and implemented gradient algorithms for noise cancellation in CR system's [15 - 16]. Therefore, this paper investigates the efficiency in using PSO for denoising signals in CR systems. The paper also compares the efficiency of PSO with that of the LMS algorithm. For the purpose of evaluating the performance of each algorithm, simulations are designed to model data-transmission between two cognitive radio units. At the receiver end, both additive white Gaussian noise (AWGN) and non-linear random noise are added to the received signal to replicate the system-induced noise in cognitive radios. The adaptive filtering system in this paper is based on the system design of an adaptive line enhancer (ALE), details of which are discussed in the next section. The rest of the paper is organized as follows. In section II, the system model and the two algorithms are described. In section III, real-time signals and the results of the two algorithms are discussed and compared. Finally, conclusions are drawn and future works based on the findings in this paper are outlined in section IV.

## II. METHODOLOGY

Fig. 1 shows the general system model which includes a CR transmitter and a CR receiver. Information bits are modulated using *M*-ary phase shift keying (M-PSK) modulation scheme and converted to transmission signal $x(t)$. Analogous to the simulated model, a practical setup with two SDR units are used to observe the transmitted and received data using gigahertz and megahertz range frequency bands for transmitting M-PSK modulated signals. For the simulations and similar to the real-time signals, the simulated transmitted signal $x(t)$ goes through a medium, where noise $n(t)$ is added to form the received signal $r(t)$.

At the receiver end, noisy signal $r(t)$ is then sampled and forwarded to adaptive noise cancellation block. An ALE based filtering system is followed for noise cancellation. Unlike active noise control (ANC) filtering systems that require a primary and reference sensor, ALE uses a single sensor. Received sampled noisy signal, $d[n]$ is fed to the ALE, which introduces a delay $Z^{-\Delta}$ to produce the delayed version of $d[n]$, denoted as $\hat{y}[n]$, as shown in Fig. 2. Output $y[n]$ is the noise free received signal estimated by updating the weight coefficients $W[n]$ of the filter and can be expressed as:

$$y[n] = \hat{Y}[n]W[n] \qquad (1)$$
$$\hat{Y}[n] = (\hat{y}[n], \hat{y}[n-1], \dots, \hat{y}[n-L+1]) \qquad (2)$$
$$W[n] = [W_1, W_2, \dots, W_L]^T \qquad (3)$$

where, $L$ is the adaptive filter order and $T$ indicates the transpose of the vector. Optimal weight is found when the error signal $e[n]$, which is the difference between the received samples $d[n]$ and output $y[n]$, is minimized.

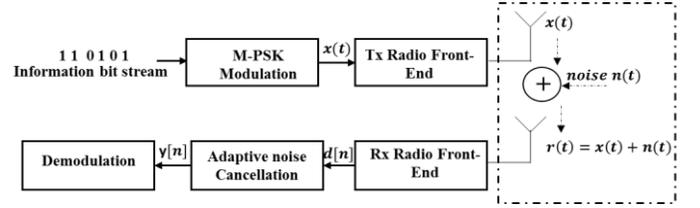
Fig. 1. System model with the communication blocks exchanging signals in passband

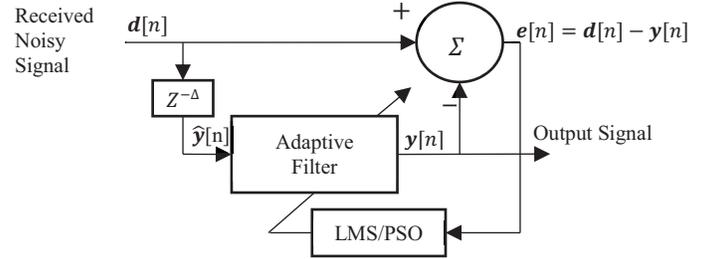
Fig. 2. Baseband block diagram showing the adaptive line enhancer based filtering system

The error signal can be written as:

$$e[n] = d[n] - y[n] \qquad (4)$$

The filtered output is then reconstructed and processed by the analog-to-digital converter to baseband received bits using the demodulation block.

### A. Adaptive noise cancellation using PSO

PSO is one of the evolutionary algorithms that is based on stochastic global optimization technique [11, 18]. For adaptive noise cancellation, PSO's objective is to minimize the residual noise by locating optimal weight coefficients for the adaptive filter. Precisely, a cost function is defined which is calculated by estimating the mean square error (MSE) between the received samples $d[n]$ and the adaptive filter output $y[n]$. The cost function can be written as:

$$C_{i,k} = \frac{1}{H}\sum_{n=1}^{H} e_{i,k}[n]^2, \qquad (5)$$

where $e_{i,k}[n]$ is the error signal at $k^{th}$ iteration for $i^{th}$ particle and $H$ is the number of input samples to the filter. As in (1), the output $y[n]$ is the result of updating $\hat{y}[n]$ with the weight coefficients supplied by PSO algorithm to the adaptive filter. PSO initializes set of particles, defining each particle's position and an initial velocity of zero. The position vector represents the weight coefficients, initialized as $N$ number of random solutions $w_i = [w_1, w_2, \dots, w_L]$ where, $i = 1, \dots N$. With the first set of particle positions, values of the cost function $C_{i,k}$ are evaluated for $N$ particles and $k$ iterations. Respective particle position for the minimum value of cost function is set as $P_{bestcost}$. Velocity of $N$ particles for $k$ iterations is defined as:

$$v_{i,k} = v_{i,k-1} + c_1 r_1 (P_{bestcost,k} - w_{i,k-1})$$
$$+ c_2 r_2 (P_{globalbest,k} - w_{i,k-1}) \qquad (6)$$

where, $c_1$, $c_2$ are learning coefficients, $v_{i,k}$, $w_{i,k-1}$ are the velocity and position, respectively, and $r_1, r_2$ are uniformly distributed random numbers within the range of 0 and 1. Position of the $i^{th}$ particle at $k^{th}$ iteration is updated using:

$$w_{i,k} = w_{i,k-1} + v_{i,k} \quad (7)$$

At the $k^{th}$ iteration, position $P_{bestcost}$ is the local best position and $P_{globalbest}$ is the global best position among the $k$ iterations. Until the algorithm converges to a global optimum solution or a maximum number of iteration is reached, these processes are repeated, as shown in the flowchart in Fig. 3.

*B. Adaptive noise cancellation using LMS*

LMS is a gradient descent algorithm that is initialized with an assigned value and follows the negative of gradient to reach the desired local minimum. LMS employs a step size, which can be described as the guiding factor to decide on the direction of the negative descent from one point to the next. Weight update in LMS can be expressed as:

$$W[n+1] = W[n] + \mu e[n]\widehat{Y}[n] \quad (8)$$

where, $W[n]$ is the weight vector and $\mu$ is the step size, which controls the convergence rate. To minimize the error surface or the error signal $e[n]$, it is preferred that the step size be chosen small so as to achieve the optimal convergence speed [12]. Optimal selection of the step size is one of the major performance requirements of adaptive algorithm. As in (1), output signal is then estimated with the updated filter coefficients. The flowchart for the LMS algorithm is shown in Fig. 4.

### III. RESULTS AND DISCUSSION

PSO and LMS algorithms were implemented using MATLAB as a platform. For all simulations, at the transmitter the bit stream is generated to create a signal of $H$=10,000 samples and modulated using M-PSK modulation with M=2.

At the receiver, AWGN and non-linear random noise were added to the transmitted signal and then filtered using PSO and LMS algorithms. Two metrics were used to evaluate and compare the efficiencies of two algorithms: bit error rate (BER) and mean squared error (MSE). BER is defined as the number of bits in error divided by the total number of transferred bits during a studied time interval. It is given by:

$$BER = \frac{Number\ of\ Corrupted\ Bits}{Total\ Number\ of\ Transmitted\ Bits} \quad (9)$$

MSE represents the average of the squares of the errors or deviations, that is, the difference between the noisy signal and the output of the filter. It is defined as:

$$MSE = \sum_{l=1}^{H}(Noisy\ Signal - Filter\ Output)^2/H \quad (10)$$

where, $H$ is the length of the received signal.

The real-time signals observed from the practical setup of two SDR units and examples of results corresponding to MSE and BER for both PSO and LMS are shown in Fig. 5 through 11.

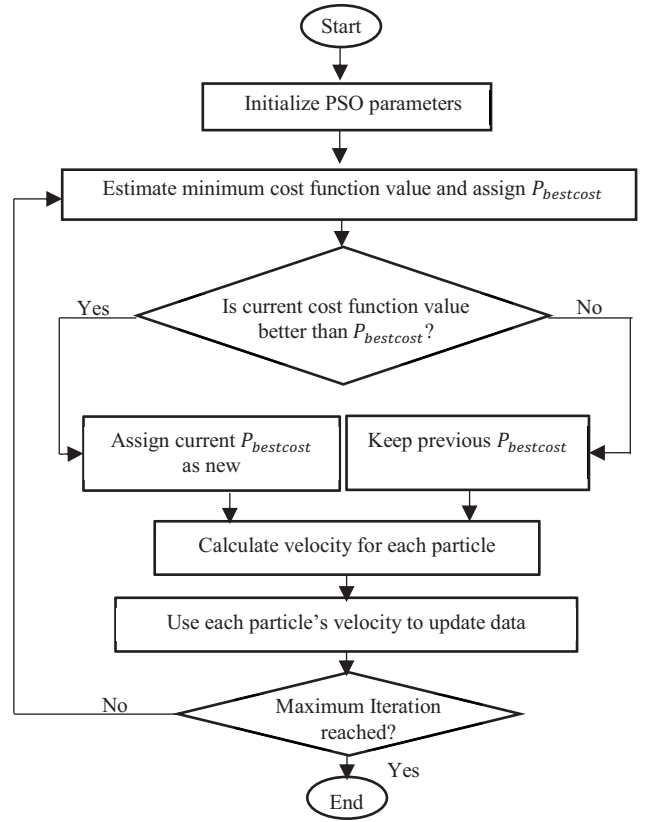

Fig. 3. Flowchart of PSO algorithm

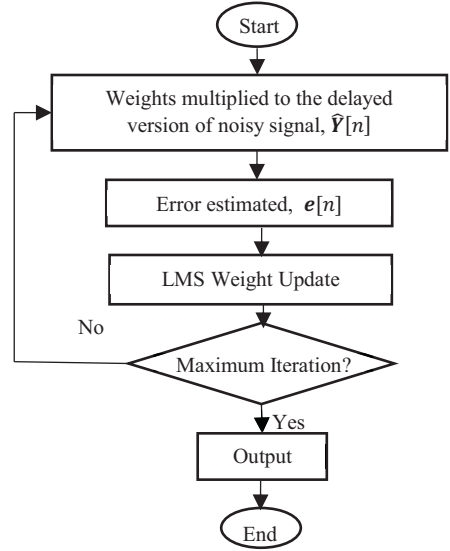

Fig. 4. Flowchart of Least Mean Square algorithm

As stated previously, one of the major drawbacks of LMS algorithm is its degraded performance with non-linear signals. Thus, we performed similar simulations for signals distorted with both AWGN and non-linear random noise. Moreover, to acknowledge CR's dynamic spectrum access capabilities, performance of these algorithms are investigated for multiple frequency ranges: 2.4GHz, 5.8 GHz and 60 MHz covering both licensed and unlicensed frequency bands used by CR systems.

In addition to simulating M-PSK modulated signals for the different frequency ranges, real time BPSK modulated signals using software defined radio (SDR) units are also investigated to observe the impact of additional noise sources in CR systems. Fig. 5 shows the practical setup using USRP N200 SDR units from Ettus Research. At this phase of the presented work, a full-duplex communication between SDR units could not be established and therefore, noise saturation due to the co-located CR transmitter and receiver antennas operating at the same time and frequency is not investigated. . Hence, the real-time signals are only used for observation and both the AWGN and non-linear noisy signal are simulated to maintain similar simulation environment and control parameters for all cases

Fig. 6 shows the power spectrum as the function of frequency for a) USRP-1 transmitted signal, b) USRP-2 received noisy signal, c) simulated transmitted signal, and d) simulated received noisy signal. It is observed that the simulated received signal corrupted by AWGN and random non-linear noise is similar to the real-time received noisy signal. However, the simulated non-linear signal was generated to have some noise induced additional spectrum at lower frequency ranges to replicate the noise saturation in co-located CR antennas. As can be seen in Fig. 6d, the spikes of low frequency ranges represent the non-linear noise induced additional spectrum besides the 2.4GHz band.

Fig. 7 illustrates the impact of particle size on the convergence characteristic of PSO. The results were obtained with added Gaussian noise, a filter order L=5, and a fixed signal-to-noise ratio (SNR) at -2 dB. As shown in the figure, global best costs of PSO algorithm for 6 different particle sizes are updated after every iterations. For each particle size the algorithm was run for 60 iterations and it is observed that particle size of 60 converges within 10 to 15 iterations whereas rest of the particle sizes converge around the 22$^{nd}$ iteration. As can be seen from this Fig. 7, global best costs do not vary significantly. Therefore, a particle size of 60 is chosen for the rest of the simulations considering a number of factors, such as convergence, complexity and processing time.

Fig. 8 illustrates the effect of different step size values on the corresponding MSE for the LMS filtered output of AWGN corrupted signal. These results are obtained for a signal with a fixed SNR of -2 dB and a filter order of L=5 as LMS tends to operate well with smaller filter orders [12]. According to the result, for step sizes less than 0.01, MSE values are high. Between step sizes 0.01 and 0.04, MSE values were found to be the lowest. However, after step size of 0.04, MSE starts increasing again. This increase is due to the fact that with large step size values, LMS algorithm fails to find the optimal weight coefficients to minimize the error. Therefore, large step size values are considered inappropriate and direct the LMS algorithm to diverge and increase the MSE.

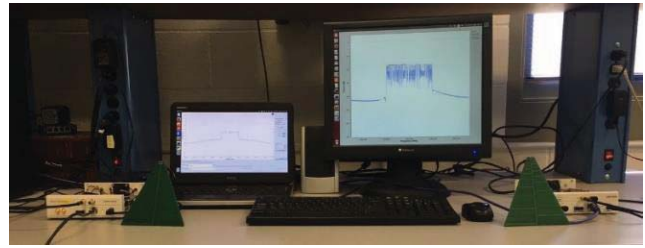

Fig. 5. Practical setup of USRP N200 Series from Ettus Research

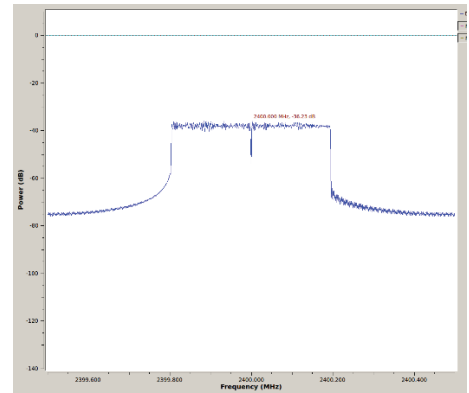

Fig. 6a. Transmitted signal from USRP -1

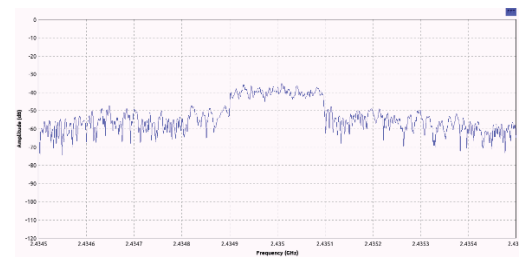

Fig. 6b. Received noisy signal by USRP-2

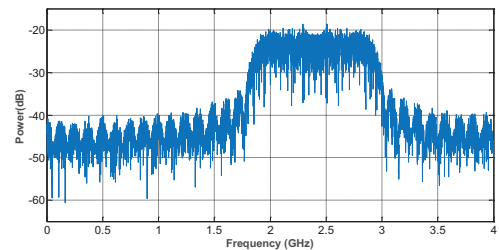

Fig. 6c. Simulated 2.4GHz transmitted signal

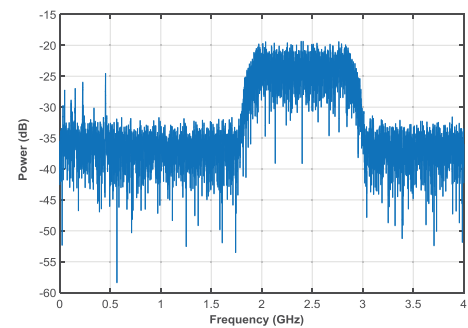

Fig. 6d. Simulated received signal with AWGN and non-linear noise

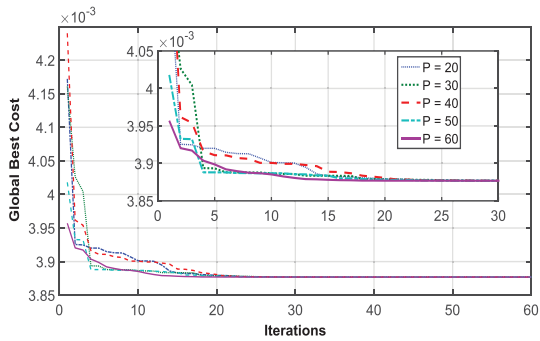

Fig. 7. Impact of particle size on PSO convergence characteristics

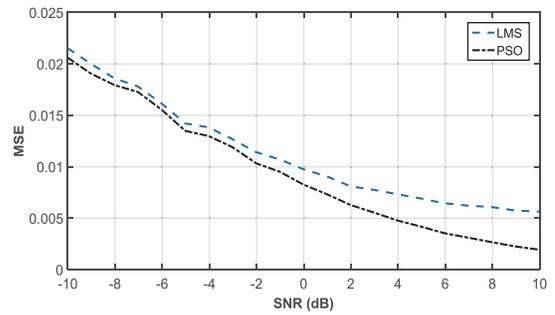

Fig. 9. MSE of PSO and LMS for varying SNR conditions.

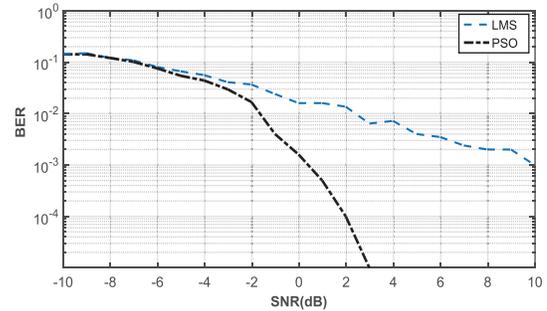

Fig. 10. BER of PSO and LMS for AWGN based noisy signal.

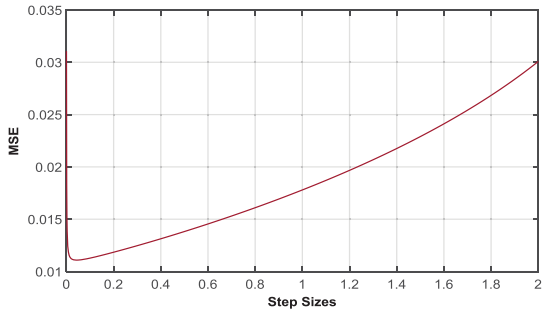

Fig. 8. Mean square error for different LMS step sizes.

Fig. 9 compares the MSE of PSO and LMS for varying SNR conditions. These results are obtained using the optimal particle size of 60 for PSO and step size of 0.01 for LMS found in previous simulations. From the figure, it is observed that MSE for both algorithms decreases as SNR increases. However, after SNR of -2 dB, the difference in MSE between PSO and LMS increases indicating better performance of the PSO than the LMS.

Performance of PSO is more notably recognized in the bit error rate analysis. Fig. 10 and 11 show performance of PSO and LMS in filtering a signal with AWGN and non-linear random noise in simulated conditions. The two figures correspond to BER for SNR conditions ranging from -10 to 10 dB. Fig. 10 compares the BER of PSO and LMS for an only AWGN added noisy signal transmitted with the 2.4GHz frequency band. The simulations were performed using a step size 0.01, particle size of 60 and filter order L=5. As one can see, BER for both the algorithms decreases at a similar rate till SNR of -6 dB. After SNR of -6 dB, the difference in BER between both algorithms increases as PSO perform significantly better than LMS. At 0.5 dB SNR, PSO achieves a bit error rate of 0.001 while LMS is found to achieve the same bit error at 10 dB SNR. Unlike LMS, PSO is able to locate the global optimum solution of an error surface and is, therefore, seen to perform better than LMS.

Fig. 11 shows the results of BER for PSO and LMS filtered non-linear signals of 2.4, 5.8 GHz and 60 MHz bands, under varying SNR conditions. As expected, under low SNR conditions, LMS and PSO perform to achieve similar BER for all the three signals. As SNR increases, PSO outperforms LMS because the effect of AWGN is decreased and the presence of non-linear random noise becomes more prominent in the received signal.

Precisely, PSO achieves the lowest BER at 10dB SNR when filtering the signal of 60 MHz. However, after an SNR of 2 dB, PSO has the minimum BER for the 2.4GHz signal and continues to have that decreasing rate till 9dB after which the 60 MHz signal outperforms. Overall the performance of both algorithms degrades when the signal is corrupted by both AWGN and non-linear random noise.

Tables I provides general performance comparisons of PSO and LMS in terms of convergence rate, and optimization efficiency. As shown in the table, PSO is more computationally complex than LMS. Unlike LMS that is dependent on appropriate step size determination, the convergence rate of PSO is not affected by any initialization variable. As global and local optimization techniques, optimization efficiency of PSO and LMS are also outlined in the table.

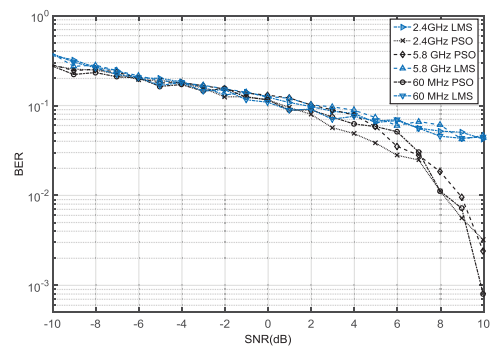

Fig. 11. BER of PSO and LMS for AWGN and non-linear random noise based received signal.

TABLE I
GENERAL COMPARISON OF PSO AND LMS PERFORMANCE

| Algorithm | Complexity | Convergence | Optimization Efficiency |
|---|---|---|---|
| PSO | Complex | Not affected by initialization variables. | Able to locate global minima |
| LMS | Simple | Affected by initialization variables, e.g. step size | Only locates local minima |

IV. CONCLUSIONS & FUTURE WORKS

This paper described the implementation of PSO and LMS algorithms. Extensive simulations were performed by modelling realistic communication systems and signals with Gaussian and non-linear random noise. The efficiencies of the two algorithms were evaluated and compared using BER and MSE as metrics. The results showed that PSO algorithm has a significantly better BER in the case of Gaussian noise compared to LMS. However, although both algorithms show degrading performance in the case of non-linear random noise, PSO still outperforms LMS. In addition, MSE of the algorithms for varying SNR conditions were discussed. The results show that MSE for PSO is lower than that of LMS with increasing SNR. The impacts of different particle sizes and step sizes on the MSE of PSO and LMS were also examined.

Motivated by the findings described in this paper, the project aims to continue the work by studying and evaluating performance of other evolutionary algorithms and by designing more practical experiments to observe the impact of noise due to changes in frequency, protocols, modulation and signal power during SDR's operation. Eventually, all the algorithms are intended to be developed as GNU Radio signal processing blocks, which will enable the use of these algorithms in practical setup of SDR networks.